\documentclass[10pt]{iopart}

\usepackage{graphicx}
\usepackage[usenames]{color}
 
\begin{document}


\title[Spatiotemporal  vortex light   bullets  in a cubic-quintic nonlinear medium]{Elastic collision  and breather formation  of  spatiotemporal  vortex light   bullets  in a cubic-quintic nonlinear medium}

\author{ S. K. Adhikari}
 
\address{
Instituto de F\'{\i}sica Te\'orica, UNESP - Universidade Estadual Paulista, 01.140-070 S\~ao Paulo, S\~ao Paulo, Brazil
} 
\ead{adhikari44@yahoo.com}
\vspace{10pt}
\begin{indented}
\item[]February 2017
\end{indented}

\begin{abstract}

The statics and dynamics  of a stable, mobile   three-dimensional (3D) spatiotemporal vortex light bullet
in  a cubic-quintic nonlinear medium with a focusing cubic nonlinearity above a critical 
value and any  defocusing quintic nonlinearity is considered. 
The present study is based on an analytic variational approximation and a full numerical solution of the 3D nonlinear 
Schr\"odinger equation. 
The 3D vortex  bullet  can  propagate with a constant velocity. 
 Stability of the vortex bullet  is established numerically and variationally.
The collision between two vortex bullets moving along the angular momentum  axis is considered.
At large velocities the collision is quasi elastic with the bullets emerging after collision with practically 
no distortion. At small velocities two bullets coalesce to form a single entity called a breather.

\end{abstract}

\noindent{\it Keywords:\/} NLS equation, vortex bullet, soliton

\pacs{05.45.-a, 42.65.Tg, 42.81.Dp}

\submitto{Laser Physics Letters}

\maketitle

 \section{Introduction}

An one-dimensional (1D)  bright soliton  with a cubic nonlinearity,  capable of moving  at a constant velocity  \cite{book,sol},    
 has been observed  in nonlinear optics \cite{book,sol}  in both 
 temporal \cite{temporal}  and spatial \cite{1996} varieties.
Although,   a three-dimensional (3D)   spatiotemporal soliton
cannot be formed with a cubic nonlinearity 
due to collapse \cite{book,collapse},  
 the soliton  can be stabilized in higher dimensions
 with  a saturable  \cite{collapse,lbth}
 or a modified nonlinearity 
\cite{quartic}, or   with a cubic-quintic nonlinearity \cite{ska} or with a modified dispersion.  
A two-dimensional (2D)   spatiotemporal optical soliton has 
been observed \cite{stempo} in a saturable nonlinearity generated by the cascading of quadratic
nonlinear processes. A 2D spatial soliton in
a cubic-quintic medium has been suggested \cite{2dcqth} and realized 
experimentally \cite{2dcqexp}. The generation of a stable 2D vortex soliton in a cubic-quintic medium has been suggested \cite{2dcqvor}.
{ There has also been a study of the dynamics of the vortex pulsed beam in a medium with nonlinearities of opposite sign \cite{2dcqvorx} and of interacting vortices in Bose-Einstein condensate (BEC) \cite{2dcqvory}.}

 A 3D  spatiotemporal optical soliton, commonly known as 
 a light bullet,  was realized experimentally in arrays of wave guides \cite{lbexp}. There are many theoretical $-$ numerical and analytical $-$  studies on light bullets  using the 3D nonlinear Schr\"odinger (NLS) equation \cite{book} with a modified nonlinearity \cite{lbth,quartic}, nonlinear dissipation \cite{gl}, and/or dispersion \cite{sadhan2}.  
Dispersion  and nonlinearity management can stabilize light bullets in a medium with cubic nonlinearity
 \cite{boris1}.  Solitons have also been studied in the coupled NLS equation \cite{cpld}.
Recently, we studied  \cite{ska} the formation of a    3D spatiotemporal
light bullet \cite{collapse,lbth} in a cubic-quintic medium  for a defocusing quintic nonlinearity and a 
focusing cubic nonlinearity. 
A cubic-quintic medium is of experimental interest also. The study with a polydiacetylene paratoluene sulfonate  crystal in the wavelength region near 1600 nm shows that 
the refractive index versus input intensity correlation leads to a cubic-quintic form of 
nonlinearity in the NLS equation \cite{book,cqexpt}. The cubic-quintic nonlinearity also 
arises in a low intensity expansion of the saturable nonlinearity used in the pioneering 
study of light bullets \cite{lbth}.

In this Letter we demonstrate the stabilization of a 3D spatiotemporal vortex (rotating)  light bullet in a cubic-quintic medium and study its statics 
and dynamics employing variational and numerical solutions of the 3D nonlinear Schr\"odinger equation. 
The vortex light  bullet   is capable of moving  without deformation  with a constant velocity. 
We  study the  collision between two  vortex light bullets moving along the spinning axis.  
Such a collision in 3D is expected  to be inelastic with loss of energy.  
In the present numerical simulation of collision between two vortex light bullets
in different parameter domains of nonlinearities and velocities three distinct scenarios are found to take place. At sufficiently large velocities  the collision is found to be quasi elastic when the two bullets emerge after collision with practically no deformation. 
  At small   velocities 
 the collision   
is inelastic and  the bullets  form a single bound entity in an excited state
 and  last for ever and execute oscillation. We call this a breather. 
In a small domain of intermediate velocities, the bullets coalesce to form a single entity, which expands indefinitely  leading to the destruction of the bullets.

We present the 3D NLS equation used in this study  in Sec. 
\ref{II}. 
In Sec. \ref{III} we present the numerical results for stationary profiles of  3D spatiotemporal vortex light bullets.  We present numerical tests of stability of the vortex light  bullet  under a small perturbation. 
The quasi-elastic nature of collision 
of two vortex bullets at large velocities and  formation  of a breather at low velocities 
 are  demonstrated by realistic simulation. 
We end with a summary of our findings in Sec. \ref{IV}.

\section{Nonlinear Schr\"odinger equation: Variational formulation}
 
\label{II}

The 3D NLS equation we describe below to study vortex soliton has application in two areas: in nonlinear optics, where the soliton is known as 
a spatiotemporal optical vortex bullet, and in BEC.    
{In nonlinear fiber optics the 3D NLS equation is  \cite{book,agarwal}
\begin{equation} \,
{\Big [}  i \frac{\partial }{\partial \mathrm{z}} +\frac{1}{2 \beta_0}\big(  \frac{\partial^2}{\partial \mathrm{x}^2}+
\frac{\partial^2}{\partial \mathrm{y}^2}  \big)+\frac{\beta_2}{2}\frac{\partial^2}{\partial \mathrm{t}^2}+ \gamma \vert A \vert^2
-\kappa  |A|^4
{\Big ]}  A(\mathrm{x,y,t})=0,
\label{eq0}
    \end{equation}
where the unit of the parameter $\gamma$ is W$^{-1}$m, that of $\kappa$ is   W$^{-2}$m$^3$,  that of the intensity $|A|^2$ is {$W $m$^{-2}$, that of the 
dispersion parameter  $\beta_2$ is ps$^2$/m, and that of the propagation constant $\beta_0$ is m$^{-1}$.
We define the diffraction length $L_{DF}\equiv \beta_0 \omega^2 $ and
dispersion length 
 $L_{DS}\equiv \tau^2/|\beta_2|$,  where $\omega$ is the width of the pulse, and $\tau$ is the 
time scale of the soliton \cite{malomed}.
Now   one  defines the following dimensionless variables \cite{agarwal}
\begin{eqnarray}\;
 x=\frac{\mathrm{x}}{\omega},\;  y=\frac{\mathrm{y}}{\omega}, \; t=\frac{\mathrm{t}}{\omega \sqrt{\beta_0 \beta_2}},\; z =\frac{\mathrm{z}}{L_{DF}},\; \nonumber \\
{ \Phi=\frac{A\sqrt{\gamma L_{DF}}}{\sqrt P_0}, \; p= {  P_0 } , \; q=\frac{\kappa P_0^2 }{\gamma^2 L_{DF}}.}\label{unit}
\end{eqnarray}
The scale $P_0$ is chosen to yield unit  norm: $\int|\Phi|^2 dx dy dt =1.$
 Using dimensionless variables one obtains  
the following  NLS equation 
with self-focusing cubic and self-defocusing 
quintic nonlinearity
    \cite{book}
\begin{eqnarray}
{\Big [}  i \frac{\partial }{\partial z}& \, +\frac{1}{2} \left( \frac{\partial^2}{\partial x^2} +\frac{\partial^2}{\partial y^2} 
 +\frac{\partial^2}{\partial t^2}\right)  +p \vert \Phi \vert^2
- q \vert \Phi \vert^4
{\Big ]}  \Phi({\bf r},z)=0,
\label{eq1x}
\end{eqnarray} 
where ${\bf r}\equiv \{x,y,t\}$,   
  $p$ and $q$ are  the coefficients of cubic and  quintic 
nonlinearities, respectively.  
In  (\ref{eq1x}) $x,y$ denote transverse extensions, $z$ the propagation distance,  and 
$t$ the time. The quintic nonlinearity of strength $q$ with a negative sign 
denote self-defocusing.
The plus sign before $|\Phi|^2$  
denotes a self-focusing cubic nonlinearity. 

For a vortex of  charge $L$ with circular symmetry in the $x-y$ plane, we can write $\Phi({\bf r},z)= \phi_L(\rho,t,z) \exp(i L \theta), \rho=\sqrt{x^2+y^2}, x=\rho \sin \theta, y=\rho\cos \theta $, where the function $\phi_L(\rho,t,z)$ is real with the property $\lim_{\rho \to 0} \phi_L(\rho,t,z) \to \rho^L.$  This generates an optical pulse with a dark spot at the center ($\rho=0$) and is called an optical vortex \cite{torner}. The wave function   $\Phi({\bf r},z)$ is periodic in $\theta$ with a period $2\pi$ (rotational symmetry).
Consequently, recalling 
\begin{equation}
\frac{\partial^2}{\partial x^2} +\frac{\partial^2}{\partial y^2} = \frac{\partial^2}{\partial \rho^2}+\frac{1}{\rho} \frac{\partial}{\partial \rho}+
\frac{1}{\rho^2}\frac{\partial^2}{\partial \theta^2},
\end{equation}
for unit charge $L=1$,  (\ref{eq1x}) becomes \cite{torner}
\begin{eqnarray}
{\Big [}  i \frac{\partial }{\partial z}& \, +\frac{1}{2} \left( \frac{\partial^2}{\partial \rho^2} +\frac{1}{\rho}\frac{\partial}{\partial \rho} 
 +\frac{\partial^2}{\partial t^2}\right)- \frac{1}{2\rho^2}\nonumber \\ & \, +p \vert \phi \vert^2
- q \vert \phi \vert^4
{\Big ]}  \phi(\rho,t,z)=0,
\label{eq1}
\end{eqnarray}  
where we have dropped the $L=1$ index from the wave function.

{To estimate the order of magnitude of different variables, we consider  an infrared beam of wave length 
$\lambda = 1$ $\mu$m in  a nonlinear medium of  $\beta_2= 10^{-2}$ ps$^2$/m, with the  time scale $\tau=60$ fs.  Then
the beam width $\omega \approx 239$ $\mu$m and 
 the dispersion length $L_{DS}=36$ cm. These numbers are quite similar to those in an
experiment on spatiotemporal optical bullet in a planar glass wave-guide \cite{sptex}.
Here we present the results in dimensionless units,
 which can be converted to actual experimental units using the transformations (\ref{unit}).

The   analytic model   (\ref{eq1}) is also applicable to the case of a vortex soliton in BEC \cite{becx}.
 In that case the mean-field Gross-Pitaevskii equation describing the BEC 
in the presence of an attractive two-body and repulsive three-body interactions 
 is given by \cite{lpl}
\begin{eqnarray}\label{eq21}
i\hbar \frac{\partial \psi({\bf r},t)}{\partial t} & =\biggr[-\frac{\hbar^2}{2m} \nabla^2_{\bf r}   -\frac{4\pi  |a| \hbar^2N}{m}|  \psi({\bf r},t) |^2 \nonumber \\
&  +\frac{\hbar N^2 K_3}{2} |  \psi({\bf r},t) |^4  \biggr]  \psi({\bf r},t),
 \end{eqnarray}
where  
$m$ is the mass of each atom of the BEC,  $\psi({\bf r},t)$ is the condensate wave function at space point ${\bf r}=\{x,y,z\}$ and time $t$,   $\rho=\sqrt{x^2+y^2}$, $a$ is the $s$-wave scattering length of atoms, $K_3$ is the three-body interaction term, and $N$ is the number of atoms. Equation (\ref{eq21}) can be written in the following dimensionless form after a redefinition of the variables
\begin{eqnarray}\label{eq3}
i\frac{\partial \psi({\bf r},t)}{\partial t}   =\biggr[-\frac { \nabla^2_{\bf r}}{2}   - p |\psi({\bf r},t) |^2  
   +q|  \psi({\bf r},t) |^4  \biggr]  \psi({\bf r},t),
 \end{eqnarray}
where 
$p=4\pi N$, $q= m N^2 K_3/(2\hbar a^4)$, length is scaled in units of $|a|$, time in $ma^2/\hbar$, $|\psi|^2$ in units of $|a|^{-3}$. Equations (\ref{eq1x})
and (\ref{eq3}) are mathematically the same, but the interpretation of the various terms in them is distinct. A BEC vortex soliton can be introduced in   
 (\ref{eq3}) in a similar fashion as in the case of optical pulse, viz.   (\ref{eq1}). 
In the following we will discuss mostly the 
case of  spatiotemporal vortex light  bullet in cubic-quintic medium. 
Nevertheless, the similarity of the mathematical models (\ref{eq1x}) and (\ref{eq3}) ensures the the possibility of generating a 3D vortex soliton in a BEC with 
repulsive three-body and attractive two-body interactions.  
  
 For an analytic understanding of the 
formation of a spinning light bullet (a vortex soliton of unit charge),
we consider the   Lagrange  variational formulation of an optical pulse \cite{var}. 
 In this axially symmetric problem, convenient analytic
 variational approximation of the vortex bullet is   \cite{var,pg}
\begin{eqnarray}\label{eq2}
 \phi(\rho,t,z)&=\frac{\pi^{-3/4}\rho}{\sigma_1^2(z)\sqrt{\sigma_2(z)}}\exp\biggr[-\frac{\rho ^2}{2\sigma_1^2(z)}- \frac{t^2}{2\sigma_2^2(z)}\nonumber \\
 &+ i \alpha(z) \rho^2+i \beta(z)   t^2\biggr],
\end{eqnarray}
where $r^2=\rho^2+t^2$, $\sigma_1(z)$ and $\sigma_2(z)$ are radial and axial  widths, respectively, and $\alpha(z), \beta(z)$ are corresponding chirps. 
 The (generalized) Lagrangian  density corresponding to   (\ref{eq1}) is given by
\begin{eqnarray}\label{eq31}
{\cal L}(\rho,t,z) &= \frac{i}{2}\left(\phi \dot \phi^*-   \phi^* {\dot \phi} \right) +\frac{|\nabla \phi(\rho,t,z) |^2}{2} +\frac{|\phi(\rho,t,z)|^2}{2\rho^2}\nonumber  \\ &-\frac{p}{2}| \phi(\rho,t,z)|^4
+\frac{q}{3}| \phi(\rho,t,z)|^6,
\end{eqnarray}
where the overhead dot denotes $z$-derivative.
Equation (\ref{eq1}) can be obtained by extremizing the functional (\ref{eq31}) \cite{var}. 
Consequently, the effective Lagrangian function $\overline L(\sigma_1,\sigma_2,\alpha,\beta) \equiv 2\pi \int {\cal L}(\rho,t,z) dt \rho d\rho$ becomes
\begin{eqnarray}\label{eq4}
\overline L(\sigma_1,\sigma_2,\alpha,\beta) &= \sigma_2^2\biggr(\frac{\dot \beta}{2}+\beta^2 \biggr)+ 
 2\sigma_1^2({\dot \alpha}+2\alpha^2)+\frac{1}{\sigma_1^2}\nonumber \\ &+\frac{1}{4\sigma_2^2}
-\frac{p\pi^{-3/2}}{8\sqrt 2  \sigma_1^2\sigma_2}
+\frac{2q\pi^{-3}}{81\sqrt 3  \sigma_1^4 \sigma_2^2}.
\end{eqnarray}   
The variational parameters $\nu\equiv \sigma_1, \sigma_2, \alpha, \beta $ are obtained from the Euler-Lagrangian equations
\begin{equation}
\frac{d}{dz} \frac{\partial \overline L}{\partial \dot \nu}=  \frac{\partial \overline L}{\partial  \nu}
\end{equation}
After some straightforward algebra the four Euler-Lagrangian equations lead to the following dynamical equations for the widths:
  \begin{eqnarray}\label{xx}
\frac{1}{\sigma_1^3}&- \frac{p\pi^{-3/2}}{8\sqrt 2 \sigma_1^3 \sigma_ 2}+ \frac{4q \pi^{-3}}{81\sqrt 3 \sigma_1^5 \sigma_2^2}=  \ddot \sigma_1,\\
\frac{1}{\sigma_2^3}&- \frac{p\pi^{-3/2}}{4\sqrt 2 \sigma_1^2 \sigma_ 2^2}+ \frac{8q \pi^{-3}}{81\sqrt 3 \sigma_1^4 \sigma_2^3}=\ddot\sigma_2.\label{yy}
 \end{eqnarray}
   The stationary profile of the vortex bullet is obtained by setting the $z$-derivatives on the right-hand sides of   (\ref{xx})-(\ref{yy}) \cite{var}:
  \begin{eqnarray}\label{xxx}
\frac{1}{\sigma_1^3}&- \frac{p\pi^{-3/2}}{8\sqrt 2 \sigma_1^3 \sigma_ 2}+ \frac{4q \pi^{-3}}{81\sqrt 3 \sigma_1^5 \sigma_2^2}=0,\\
\frac{1}{\sigma_2^3}&- \frac{p\pi^{-3/2}}{4\sqrt 2 \sigma_1^2 \sigma_ 2^2}+ \frac{8q \pi^{-3}}{81\sqrt 3 \sigma_1^4 \sigma_2^3}=0.\label{yyy}
 \end{eqnarray}
Equations   (\ref{xxx})-(\ref{yyy}) correspond to the global minimum of a {\it conserved} $\alpha$- and $\beta$-independent  effective Lagrangian 
$L(\sigma_1,\sigma_2)\equiv \overline L(\sigma_1,\sigma_2,\alpha=0,\beta=0)$: $\partial L/\partial \sigma_1= \partial L/\partial \sigma_2=0$.
The function  $L(\sigma_1,\sigma_2)$ describes the Lagrangian dynamics of the widths $\sigma_1, \sigma_2$ and is independent of the generalized velocities 
$\dot \sigma_1, \dot\sigma_2$.

\begin{figure}[!t]

\begin{center}
\includegraphics[width=.49\linewidth,clip]{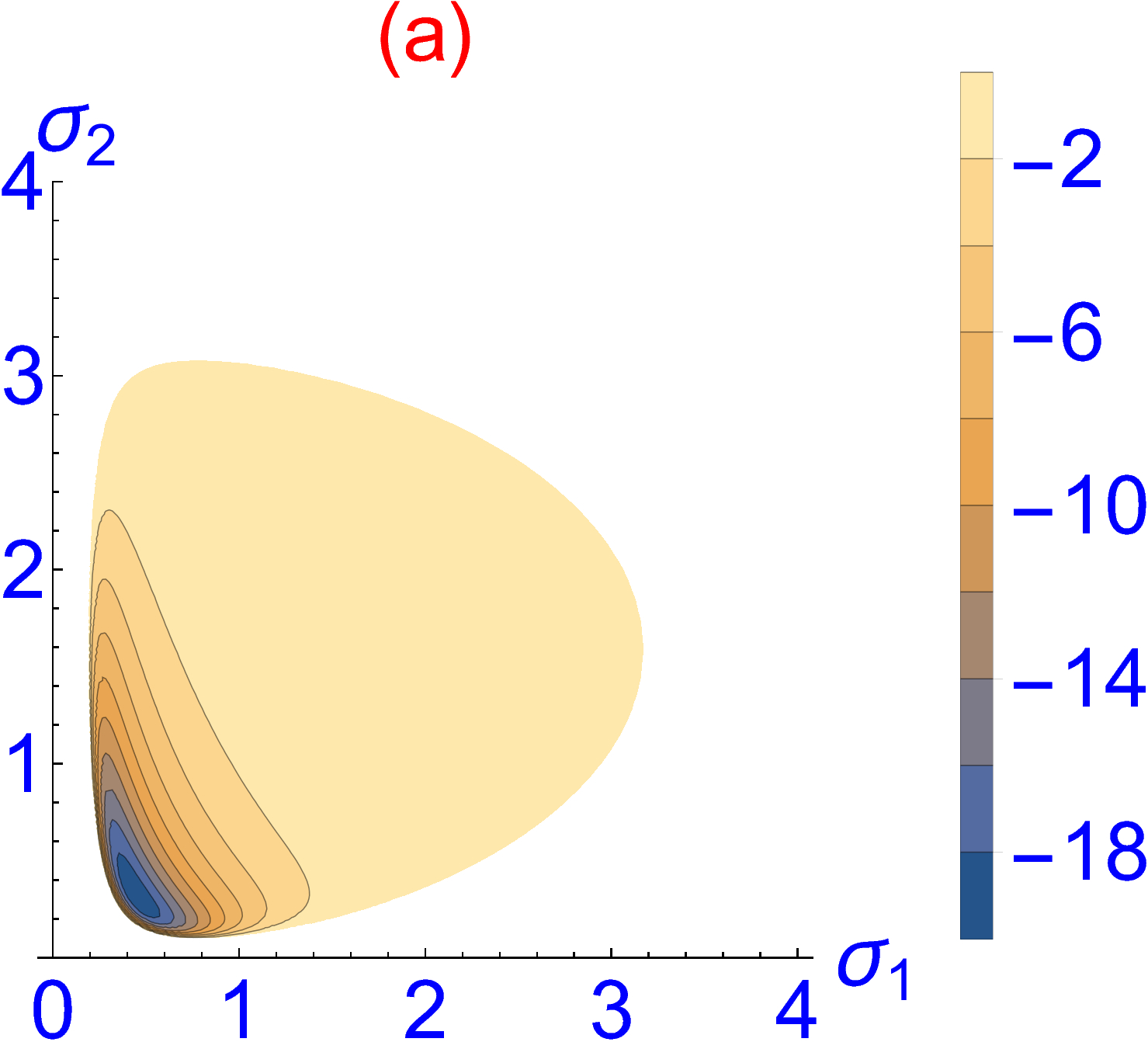} 
\includegraphics[width=.49\linewidth,clip]{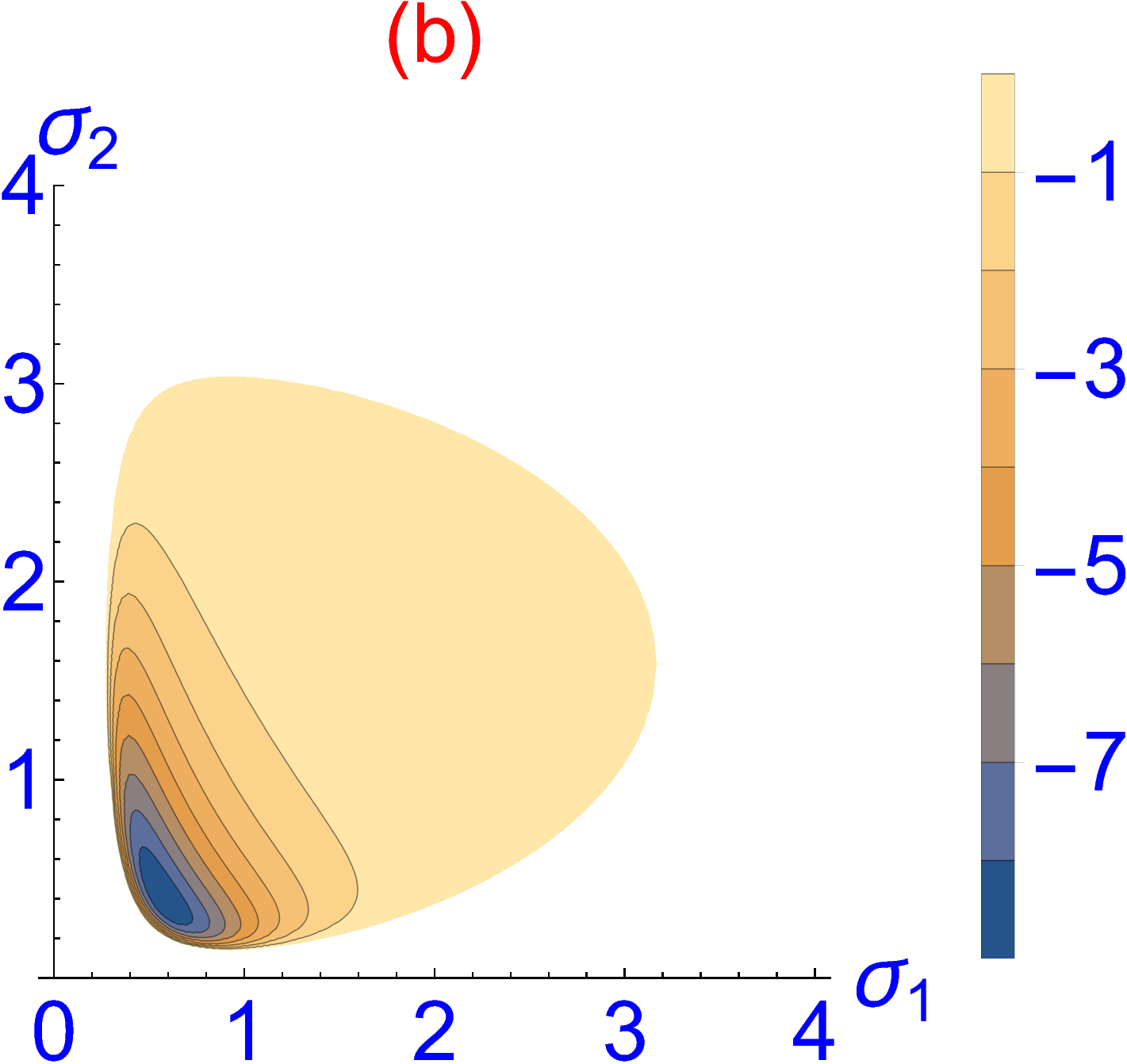}
\includegraphics[width=.49\linewidth,clip]{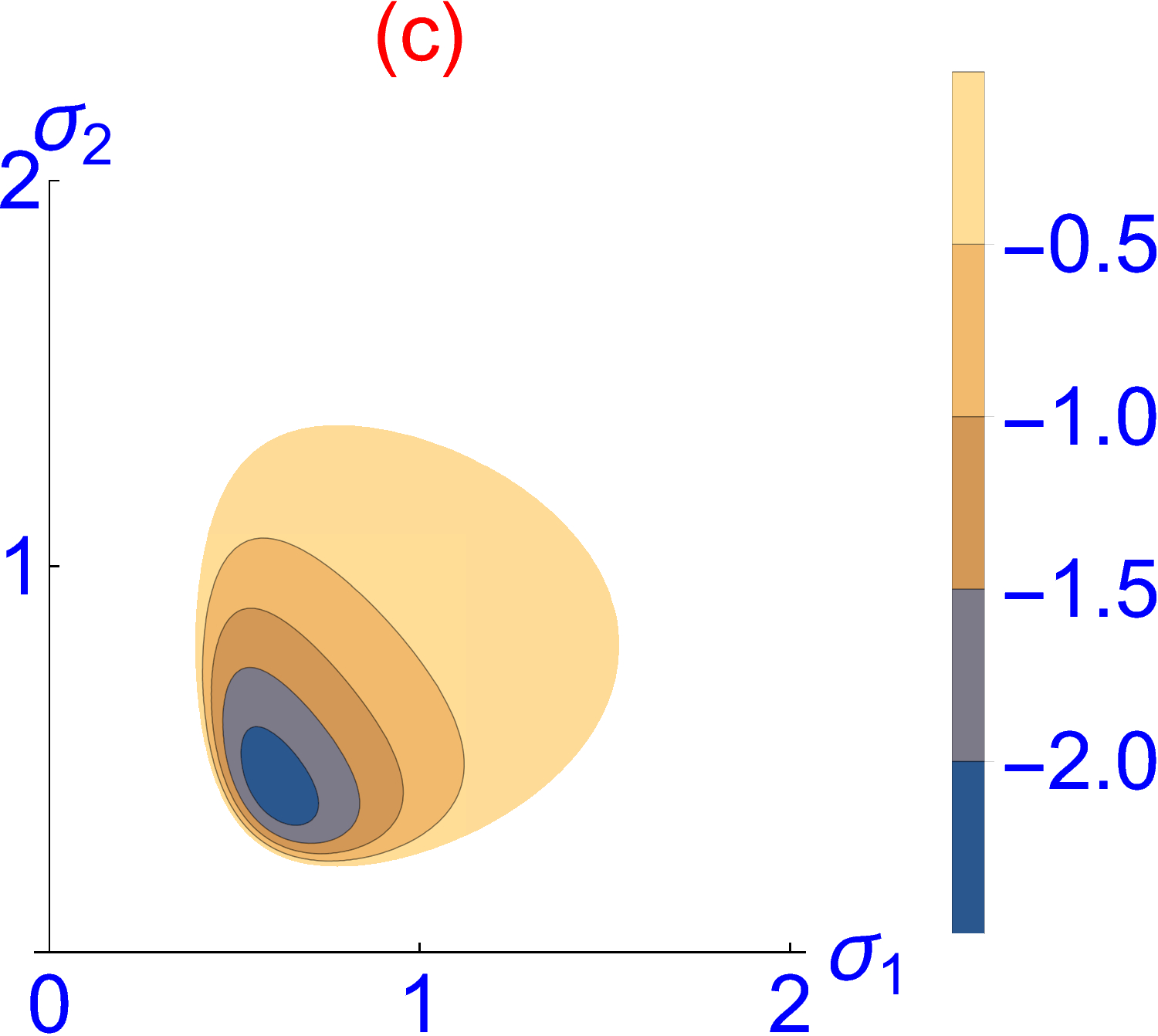}  
\includegraphics[width=.49\linewidth,clip]{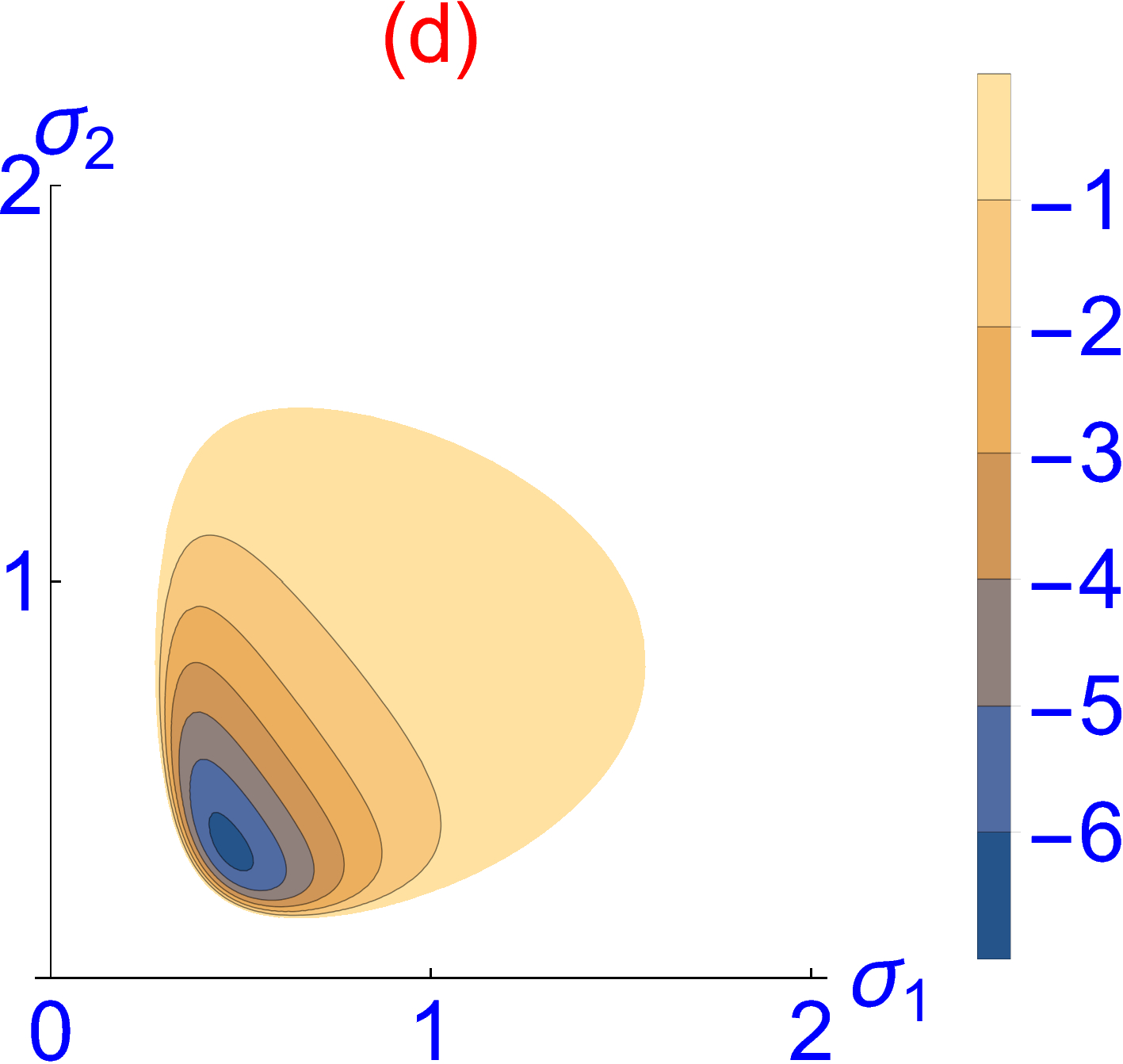}  
\caption{ (Color online) Contour plot of Lagrangian $L(\sigma_1,\sigma_2)$ (\ref{eq4}) as a function of $\sigma_1$ and $\sigma_2$ for  (a) $p=200, q=200$, (b)  $p=200, q=400$, (c) $p=100, q=200$, (d) $p=100, q=100$. The Lagrangian is negative in the shaded region and positive outside. 
}\label{figure1} \end{center}

\end{figure}

\begin{figure}[!t]

\begin{center}

\includegraphics[width=.49\linewidth,clip]{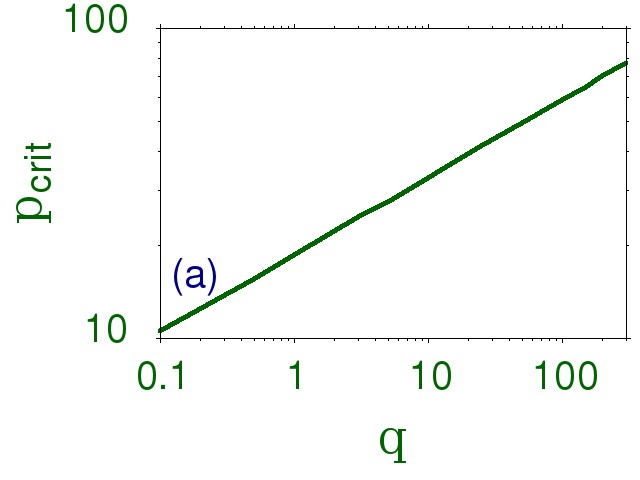}
\includegraphics[width=.49\linewidth,clip]{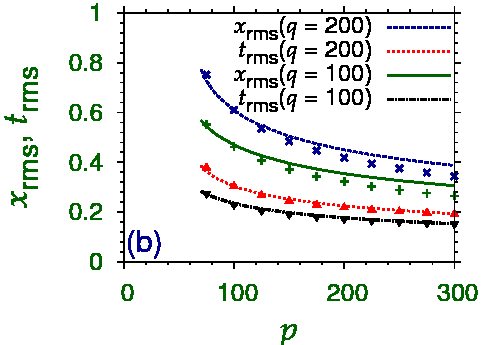}
\includegraphics[width=.49\linewidth,clip]{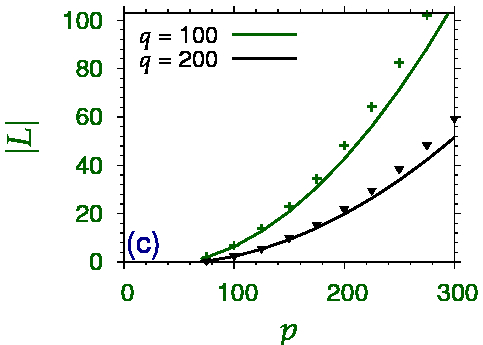}

\caption{ (Color online) (a) Variational results for $p_{\mathrm{crit}}$ versus $q$. A vortex bullet with negative Lagrangian can be formed for $p>   p_{\mathrm{crit}}$. 
(b)    Variational (line) and numerical (points) results for      rms sizes $x_{\mathrm{rms}}$ and  $t_{\mathrm{rms}}$ versus cubic nonlinearity coefficient $p$ for quintic nonlinearity coefficient $q=100$ and $200$.
(c) Variational (line) and numerical (points) results for Lagrangian   $L$  versus cubic nonlinearity coefficient $p$  for quintic nonlinearity coefficient $q=100$ and $200$.
}\label{figure2} \end{center}

\end{figure}

\section{Numerical Results}

\label{III}

The 3D NLS equation (\ref{eq1})
is generally solved by   the split-step Crank-Nicolson \cite{CPC} and Fourier spectral \cite{jpb}
methods.
 The split-step 
Crank-Nicolson method   
  in Cartesian coordinates is employed in the present study.  We use    
  $\bf r$ $=\{x,y,t\}$  step of  $0.05 \sim 0.016$,  a $z$ step of  $ 0.0005 \sim 0.0000025$ \cite{CPC} and  the number of ${\bf r}$ discretization points $128 \sim 320$. 
There are different C and FORTRAN programs for solving the NLS-type equations \cite{CPC,CPC1}
and one should use the appropriate one. 
{We use both imaginary- and real-$z$ propagation \cite{CPC} for numerical solution of the 3D NLS equation. The imaginary-$z$ propagation is appropriate  to find the stationary state and the real-$z$ propagation for the dynamics.     } In the imaginary-$z$ propagation the initial  state was taken as  in  (\ref{eq2}).

A stable bullet corresponds to a global minimum of  the conserved effective Lagrangian 
$L(\sigma_1,\sigma_2)$  at a negative value. 
To demonstrate the appearance of a global minimum, we show in figures \ref{figure1} the two-dimensional contour 
plot of the Lagrangian $L(\sigma_1,\sigma_2)$ in the  $\sigma_1-\sigma_2$  plane for
(a) $p=q=200$, (b)   $p=200,q=400$, (c) $p=100,q=200$, and  (d) $p=q=100$, where we illustrate the region with negative Lagrangian; the Lagrangian is positive 
outside this region.   The Lagrangian $L(\sigma_1,\sigma_2) $ goes to zero as $\sigma_1, \sigma_2 \to \infty$. At the origin, $\sigma_1, \sigma_2 \to 0$, and the Lagrangian $L (\sigma_1,\sigma_2) \to \infty$, which guarantees the absence of a collapsed state at the origin. 
The repulsive quintic nonlinearity   contributes 
positively to Lagrangian; so does the first two terms on the right-hand-side of  (\ref{eq4}). To make the Lagrangian (\ref{eq4}) negative, 
the cubic nonlinearity coefficient $p$ has to be larger than a   critical value, e.g. 
 $p>p_{\mathrm{crit}}$, when  
the minimum of Lagrangian could be negative corresponding to a stable vortex bullet. For  $p<p_{\mathrm{crit}}$ the  optical pulse is
too repulsive to form a vortex bullet. 
In figure \ref{figure2}(a) we show the 
variational values for $p_{\mathrm{crit}}$ versus $q$.    
We compare the numerical and variational results  for  the root-mean-square (rms) sizes $x_{\mathrm{rms}}$ and $t_{\mathrm{rms}}$ in figure \ref{figure2}(b) 
  and Lagrangian $|L|$ in figure \ref{figure2}(c), 
for $q=100$ and 200.    {The variational  Lagrangian is calculated using   (\ref{eq31})
with the numerically obtained $\sigma_1, \sigma_2 $, corresponding to the minimum of Lagrangian (\ref{eq4}) given by  (\ref{xx}) and (\ref{yy}).} 
 The variational $x_{\mathrm{rms}}\equiv \sigma_1$. The rms sizes $x_{\mathrm{rms}}, t_{\mathrm{rms}}$  and Lagrangian $L$ of figure \ref{figure2}(b)-(c), respectively,  increase with an increase of $q$ value corresponding to repulsion and decrease with an increase of $p$ value corresponding to attraction.

\begin{figure}[!t]

\begin{center}
 \includegraphics[width=.49\linewidth,clip]{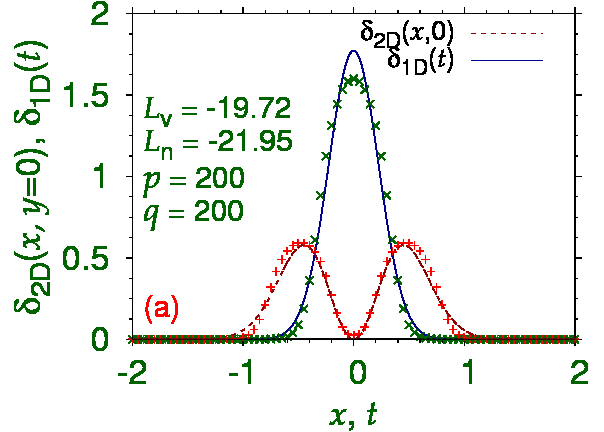}
 \includegraphics[width=.49\linewidth,clip]{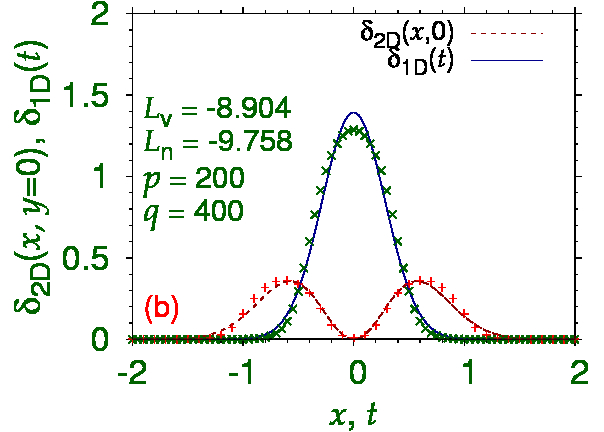}
 \includegraphics[width=.49\linewidth,clip]{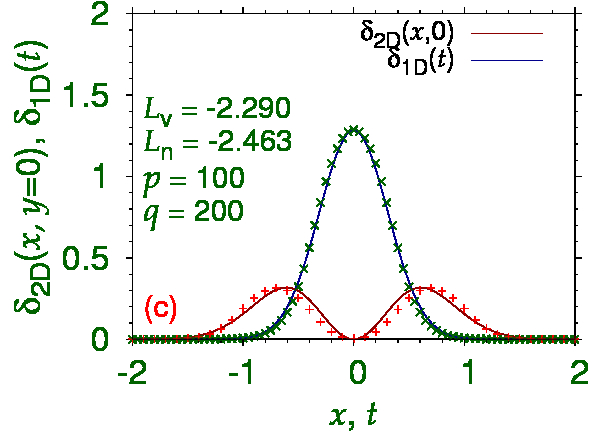}
 \includegraphics[width=.49\linewidth,clip]{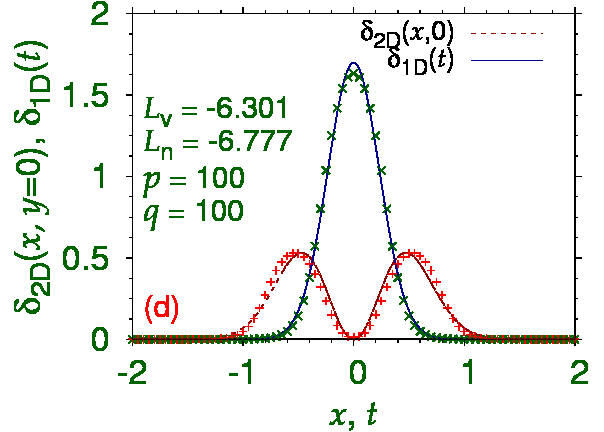}

\caption{(Color online) Numerical (line) and  variational (points) reduced  densities $\delta_{1D}(x)$ and $\delta_{2D}(x,y=0)$ for different cubic nonlinearity coefficient $p$ and quintic nonlinearity coefficient $q$: (a) $p=q=200$, (b) $p=200, q=400$, (c) $p=100, q=200$ and (d) $p=q=100.$ 
}\label{figure3} 

\end{center}

\end{figure}

To study the density distribution of the spatiotemporal vortex light bullets we define the 
reduced 1D and 2D densities  by 
\begin{eqnarray}
\delta_{\mathrm{1D}}(x) &= \int dt dy |\phi({\bf r})|^2, \\
\delta_{\mathrm{2D}}(x,y)&= \int dt |\phi({\bf r})|^2.
\end{eqnarray} 
In figure \ref{figure3} we show these reduced  densities   $\delta_{\mathrm{1D}}(x)$ and $\delta_{\mathrm{2D}}(x,0)$
as obtained from  numerical and variational
calculations for different cubic nonlinearity  coefficient $p$ and quintic nonlinearity coefficient $q$. The corresponding Lagrangian values are 
also exhibited. 
For a fixed  defocusing quintic nonlinearity coefficient
$q$ $(=200)$, the vortex bullet  is more compact with the increase 
of focusing nonlinearity coefficient $p$ resulting in more attraction as can be found in figures \ref{figure3}(a) and (c).  
For a fixed  focusing nonlinearity coefficient
$p$ = 200 and 100, the light bullet is more compact with the decrease  
of defocusing nonlinearity coefficient $q$ resulting in less repulsion as found in figures \ref{figure3}(a)-(b) and in 
 figures \ref{figure3}(c)-(d), respectively. 

 \begin{figure}[!t]

\begin{center}

\includegraphics[width=.32\linewidth,clip]{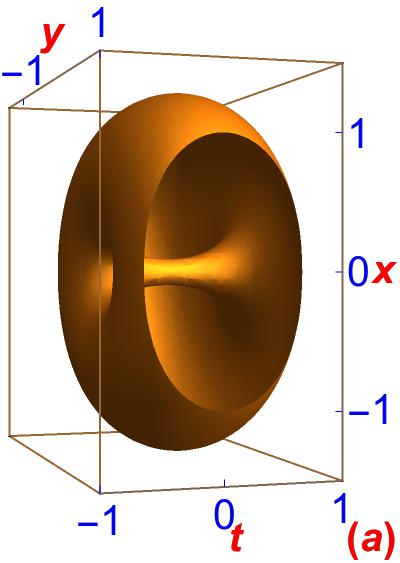}
\includegraphics[width=.32\linewidth,clip]{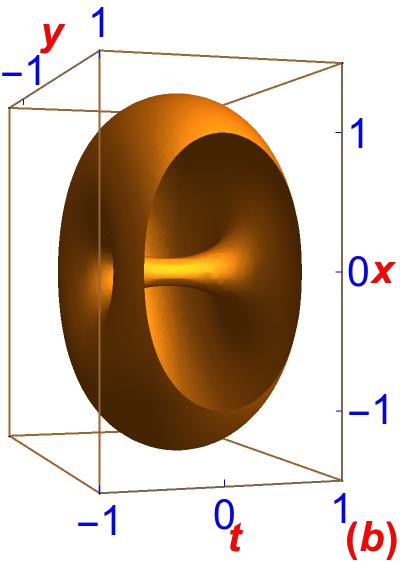}
\includegraphics[width=.32\linewidth,clip]{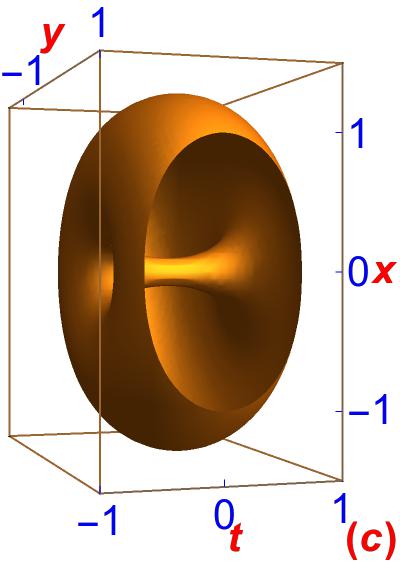}
\includegraphics[width=.9\linewidth,clip]{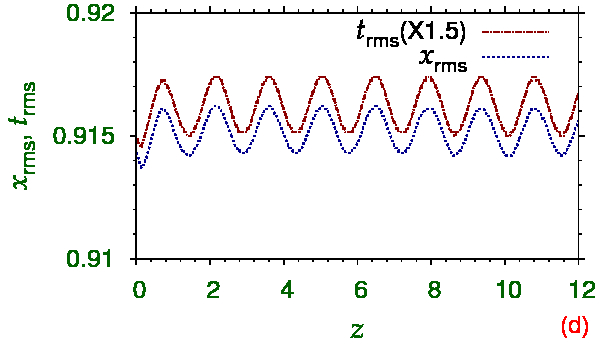}
 
\caption{ (Color online)
  Real-$z$ evolution of the vortex bullet with $p=100$ and $q=200$  by isodensity contour plot of 
$|\phi({\bf r},z)|^2$ at $(a) z=0, (b) 6, (c) 12$. The dimensionless density on the contour is $0.02$. 
(d) The rms sizes $x_{\mathrm{rms}}$ and $t_{\mathrm{rms}}$ versus $z$ during real-$z$ evolution.
}\label{figure4} \end{center}

\end{figure}

The negative-Lagrangian finite well of figure \ref{figure1} trapping the spatiotemporal vortex light bullet guarantees its stability because of Lagrangian conservation. 
Now we present a numerical test of stability of a vortex bullet. For this purpose we consider the vortex bullet shown in figure \ref{figure3}(c)
with $p=100$ and $q=200$ as calculated by imaginary-$z$ propagation. Using the imaginary-$z$
profile as the initial state we perform numerical simulation by  real-$z$ propagation.  The  real-$z$ propagation shows steady breathing oscillation of the 
vortex bullet for a large $z$ propagation.
In figures \ref{figure4}(a), (b), and (c) we show the 3D isodensity contour of the vortex bullet at $z=0$, 6 and 12. The vortex core remains intact in this propagation.  No transverse instability \cite{book} of the vortex core was found.  
 In figure \ref{figure4}(d)  we show the steady (monopole breathing) oscillation in the root-mean-square (rms) 
$x$ and $t$ sizes $x_{\mathrm{rms}}$ and $t_{\mathrm{rms}}$ versus propagation distance $z$ during real-$z$ propagation. 
The steady continued oscillation of the vortex bullet over a long distance of propagation establishes the stability of the bullet. The real-$z$ simulation was performed in full 3D
space without assuming spherical symmetry to guaranty the stability in full 3D Cartesian space.

The collision between two analytic  1D solitons is truly elastic \cite{book}  due to the conservation laws (of energy, momentum)   and such solitons pass through each other without deformation at any incident velocity.
The collision between  two 3D spatiotemporal vortex bullets is expected to be inelastic 
in general due to loss of kinetic energy leading to their  deformation. 
However, under ideal condition of large velocities the  collision between two spatiotemporal vortex bullets  can be quasi-elastic.  Under these conditions, the kinetic energy of the colliding vortex bullets is much larger than the internal  interaction energies and the duration of encounter in $z$ is small. On the other extreme when the kinetic energies of the colliding vortex bullets  are much smaller than the internal binding energies, the encounter is controlled solely by the internal interactions and the two vortex bullets after encounter form a bound entity after collision, called  a breather.   

To test the  nature of collision between  the present spatiotemporal vortex light bullets,
 we study the frontal  head-on collision of the same along the angular momentum axis ($t$). A moving bullet with velocity $v$ along the $t$ axis  can be generated by 
multiplying  the bullet wave function by $\exp(ivt)$ and performing real-$z$ simulation with this function.  
 The imaginary-$z$ profile of the light bullet  shown in  figure \ref{figure3}(c) with  $p=100, q=200$   is  used as the initial function in the real-$z$ simulation of collision, with two identical bullets  placed at $x=\pm 1.45$ initially at $z=0$. 
The vortex bullets are set in motion along the $t$ axis in opposite directions with velocity $v=43$. 
To illustrate the dynamics upon real-$z$ simulation, we plot    3D density contour $|\phi({\bf r},z)|^2 $  at different values of propagation distance $z$ 
in figures \ref{figure5}.   {     In this case the kinetic energy $v^2/2\approx 924$ is much larger than the Lagrangian ($|L| \approx 2.5$),  the collision is found to be quasi elastic and  considering the 3D nature of the collision the distortion of the vortex bullets after collision  is found to be insignificant.}

 \begin{figure}[!t]

\begin{center}
\includegraphics[trim = 0mm 0mm 0mm 0mm, clip,width=.32\linewidth]{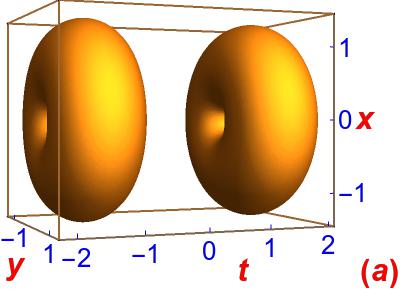}
 \includegraphics[trim = 0mm 0mm 0mm 0mm, clip,width=.32\linewidth]{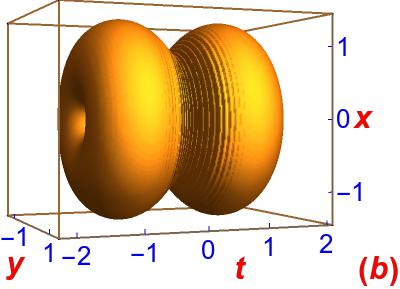} 
 \includegraphics[trim = 0mm 0mm 0mm 0mm, clip,width=.32\linewidth]{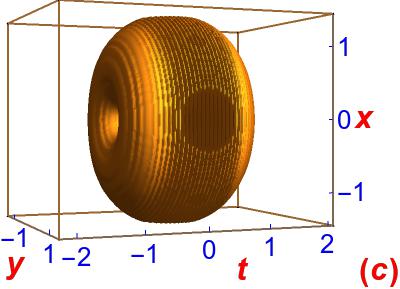}
 \includegraphics[trim = 0mm 0mm 0mm 0mm, clip,width=.32\linewidth]{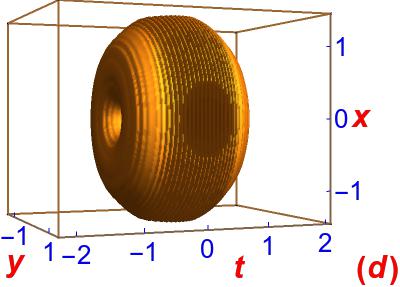}
 \includegraphics[trim = 0mm 0mm 0mm 0mm, clip,width=.32\linewidth]{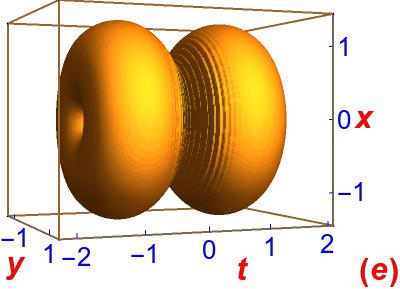}
 \includegraphics[trim = 0mm 0mm 0mm 0mm, clip,width=.32\linewidth]{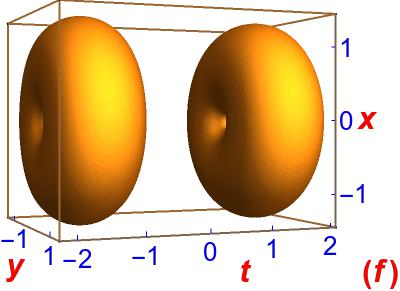}
\caption{ (Color online) Collision dynamics of two  vortex bullets , each with $p = 100,  q=200$, placed at
$t = \pm 1.45$ at $z = 0$ and set into motion in
opposite directions along the $t$ axis with the velocity
of $v=43$, illustrated by isodensity contours at  (a)
$z = 0,$ (b) = 0.0135, (c) = 0.027, (d) = 0.0405, (e) = 0.054, (f) = 0.0675. The density on the
contour is 0.02.  
}\label{figure5} \end{center}

\end{figure}

 \begin{figure}[!b]

\begin{center}
\includegraphics[trim = 0mm 0mm 5mm 0mm, clip,width=.49\linewidth]{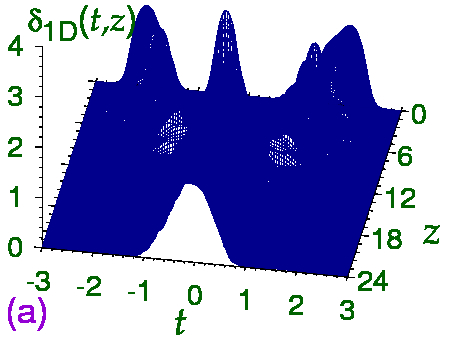}
 \includegraphics[trim = 6mm 0mm 0mm 0mm, clip,width=.49\linewidth]{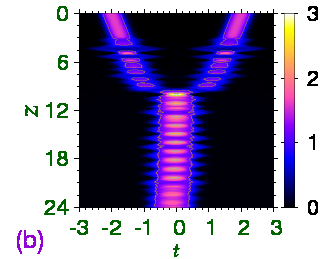} 
  
\caption{ (Color online) (a) The 1D density $\delta_{1D}(t,z) $ and (b) its 2D contour plot 
during the collision of two vortex bullets  with $p=q=200$  placed at $t=\pm 2$ 
  at $z = 0$ and set into motion in
opposite directions along the $t$ axis with the  velocity  $v=0.1$,  and the formation of a breather 
upon real-$z$ 
simulation.
}\label{figure6} \end{center}

\end{figure}

To study the inelastic collision  at very small velocities we consider two compact  bullets with 
$p=200,q=200$ and place them at $t=\pm 2$ and set them in motion with  velocity  $v=0.1$ in opposite directions along the $t$ axis.   The dynamics is illustrated by a plot of the time evolution of
1D density $\delta_{1D}(t,z)\equiv \int dx dy |\phi({\bf r},z)|^2$ versus $t,z$ in figure \ref{figure6} (a) and the corresponding contour plot is shown in figure \ref{figure6} (b).  
The two vortex bullets come close to each other at $t=0$ 
coalesce to form a breather and never separate again. The combined bound 
system remain at rest at $t=0$ 
continuing small breathing oscillation because of a small amount of
 liberated kinetic energy which creates the breather in an excited state. 
The observation of oscillating breather has been reported some
time ago in dissipative systems \cite{bm}. {  In this case the kinetic energy $v^2/2 \approx .005$ is insignificant compared to the Lagrangian ($|L| \approx 21$), and the collision 
is  fully inelastic with a destruction of individual bullets.}

 \begin{figure}[!t]

\begin{center}
\includegraphics[trim = 0mm 0mm 5mm 0mm, clip,width=.49\linewidth]{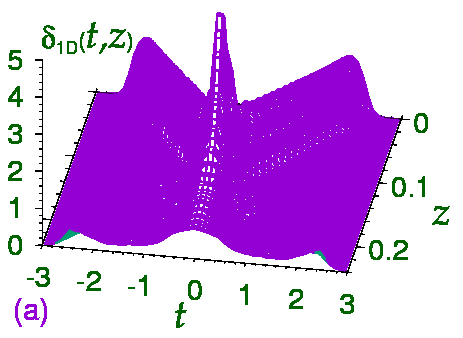}
 \includegraphics[trim = 0mm 0mm 0mm 0mm, clip,width=.49\linewidth]{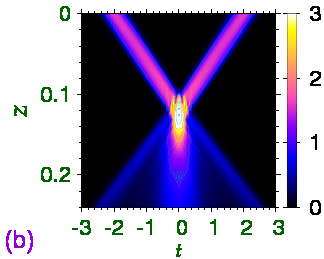} 
  
\caption{ (Color online) (a) The 1D density $\delta_{1D}(t,z) $ and (b) its 2D contour plot 
during the collision of two vortex bullets  with $p=q=200$  placed at $t=\pm 2$ 
  at $z = 0$ and set into motion in
opposite directions along the $t$ axis with the  velocity  $v=20$,  and the destruction of the 
bullets 
upon real-$z$ 
simulation. After collision the two bullets are destroyed and an expanding pulse results upon 
$z$ evolution. 
}\label{figure7} \end{center}

\end{figure}

However, as the velocity $v$ is reduced from $v\approx 40$ (elastic collision scenario presented in figure \ref{figure5}) to $v\approx 0.1$ 
(breather formation as in figures \ref{figure6}), a distortion of the vortex bullets take place after collision with eventual destruction of the vortex bullets. 
This is illustrated in figures \ref{figure7}, where apart from the two trajectories of the vortex bullets after collision a central peak can 
be visualized at $t=0$. On further reduction of the initial velocity, the central peak at $t=0$ becomes more pronounced and 
the outer tracks less prominent. Eventually, at very small velocities only the central peak corresponding to the formation of a breather 
after collision prevails, viz. figures \ref{figure6}.

\section{Summary}

\label{IV}

To summarize, we demonstrate the formation of a stable 3D spatiotemporal vortex bullet 
with cubic-quintic nonlinearity employing a variational approximation  and full 3D numerical solution of the NLS equation. The statical properties of the bullet are studied by a 
variational approximation and a numerical imaginary-$z$ solution of the 3D NLS equation. 
 The cubic nonlinearity is  
taken as focusing Kerr type above a critical value, whereas the  quintic 
nonlinearity is defocusing.  
The dynamical properties are studied by a real-$z$ solution of the NLS equation. 
In the 3D spatiotemporal case, the
vortex light  bullet  can move with a constant velocity.   At large velocities, the collision between the two 
spatiotemporal vortex light bullets    is   quasi elastic with 
no visible deformation of the final bullets.  
At small velocities, the collision is inelastic with the formation of a breather
after collision. At medium velocities the bullets can be destroyed after collision.

  \section*{Acknowledgments} 
   
We thank the Funda\c c\~ao de Amparo 
\`a
Pesquisa do Estado de S\~ao Paulo (Brazil)
(Project:  2012/00451-0)
  and  the
Conselho Nacional de Desenvolvimento   Cient\'ifico e Tecnol\'ogico (Brazil) (Project: 303280/2014-0) for 
support.
 

%
\end{document}